# Morphology of Open Clusters NGC 1857 and Czernik 20 using Clustering Algorithms[1]

Souradeep Bhattacharya[*], Vedant Mahulkar, Samay Pandaokar, Parikshit Kishor Singh

*Birla Institute of Technology and Science, Pilani 333031, Rajasthan, India*

[*] *Email: f2012553@pilani.bits-pilani.ac.in*

**ABSTRACT**

The morphology and cluster membership of the Galactic open clusters - Czernik 20 and NGC 1857 were analyzed using two different clustering algorithms. We present the maiden use of density-based spatial clustering of applications with noise (DBSCAN) to determine open cluster morphology from spatial distribution. The region of analysis has also been spatially classified using a statistical membership determination algorithm. We utilized near infrared (NIR) data for a suitably large region around the clusters from the United Kingdom Infrared Deep Sky Survey Galactic Plane Survey star catalogue database, and also from the Two Micron All Sky Survey star catalogue database. The densest regions of the cluster morphologies (1 for Czernik 20 and 2 for NGC 1857) thus identified were analyzed with a K-band extinction map and color-magnitude diagrams (CMDs). To address significant discrepancy in known distance and reddening parameters, we carried out field decontamination of these CMDs and subsequent isochrone fitting of the cleaned CMDs to obtain reliable distance and reddening parameters for the clusters (Czernik 20: $D$ = 2900 pc; $E(J-K)$ = 0.33; NGC 1857: $D$ = 2400 pc; $E(J-K)$ = 0.18-0.19).  The isochrones were also used to convert the luminosity functions for the densest regions of Czernik 20 and NGC 1857 into mass function, to derive their slopes. Additionally, a previously unknown over-density consistent with that of a star cluster is identified in the region of analysis.

# 1. INTRODUCTION

Trumpler (1930) defined star clusters as *'Star groupings which undoubtedly form physical systems (stars situated at the same distance and probably of the same origin) and which at the same time are sufficiently rich in stars for statistical investigation.'* Majority of the stars in the Galaxy are believed to have originated from such star clusters. Hence star clusters and their surroundings act like laboratories, not only for the study of star formation and evolution but also for that of stellar structures and abundance patterns. Open clusters in particular are considered as powerful tools for such studies (Friel 1995; Frinchaboy & Majewski 2008).

The first step in estimating the physical parameters and understanding stellar evolution in any open cluster is membership determination. Many attempts have previously been made at membership determination based on proper motions, radial velocities, photometric data and their combinations (Vasilevskis et al. 1958; Cabrera-Cano & Alfaro 1990; Chen et al. 2004; Zhao et al. 2006; Gao 2014). Of all these methods, those which are kinematic in nature, incorporating proper motions or radial velocities, are believed to be more reliable. Due to instrumentation restrictions, however, the reliable kinematic data are unavailable for most distant star clusters, and hence, for the majority of open clusters in the galaxy.

Thus, photometric methods are the primary tools to determine the members of any open cluster, albeit with certain inaccuracies. The inaccuracies result because the steps to determine the members using the color magnitude diagrams (CMDs) are not straightforward since they are obviously composed of stars of different stellar populations (Piatti 2012). Additionally, the parameterization of the spatial structure of an open cluster by its radial distribution alone is inadequate (Nilakshi et al. 2002) as many of these have highly irregular shapes, often with no clear centers or circular symmetry. Although previous attempts have been made to represent the stellar distribution by sophisticated methods (Chen et al. 2004), it is desirable to develop yet more refined methods in hopes of improving upon known results.

We have identified NGC 1857 (Cuffey & Shapley 1937) and Czernik 20 (Czernik 1966) as promising open clusters for such investigations because of the uncertainties surrounding their morphologies and positions. According to Babu (1989) and Sujatha et al. (2006), Czernik 20 was actually just a redetection of NGC 1857, which was centered on RA= 80˚.025, DEC= 39˚.344. UBVRI photometric observations led them to claim that NGC 1857 has an elongated morphology, and they determined various cluster parameters by isochrone fitting. However, Czernik 20 (centered on RA= 80˚.133, DEC= 39˚.539) and NGC 1857 were identified as separate clusters by Zasowski et al. (2013). Using 2MASS near infra-red (NIR) data, they found an 8′ separation between the centers of the two clusters, and further analysis produced different cluster distances and reddening from distinctly different CMDs (Czernik 20: $D = 3100$ pc, $E(J−K) = 0.35$; NGC 1857: $D = 1400$ pc; $E(J−K) = 0.07$). 2MASS data were also employed by Kharchenko et al. 2013 to estimate distance and reddening values for both the clusters (Czernik 20: $D = 2000$ pc; $E(J−K) = 0.288$; NGC 1857: $D = 3299$ pc; $E(J−K) = 0.24$). Later, Hoq & Clemens (2015) also used the same data to determine these parameters for NGC 1857 ($D = 2470$ pc; $E(J−K) = 0.31$). The discrepancy in the parameters obtained from previous studies is quite pronounced.

CMDs of star clusters are also utilized to derive observed luminosity function (LF). Using theoretical isochrones LF can be converted into the mass function (MF). The MF of open star clusters may give clues about star formation and early evolution of star clusters (Elmegreen, 1999, 2000; Richtler, 1994). For NGC 1857, Sujatha et al. (2006) derived a mass function slope corresponding to a Salpeter slope (Bastian et al. 2010) of 1.39.

In this paper, we have attempted to represent the morphologies of open clusters Czernik 20 and NGC 1857, from NIR data for the region using two different clustering algorithms. The Density-based spatial clustering of applications with noise (DBSCAN; Ester et al. 1996) is employed to spatially classify the cluster members with hard clustering, i.e., by classifying into members and non-members without assigning any kind of membership probabilities. It can find arbitrarily shaped clusters in parameter spaces that are affected by background noise. In astronomy, it has previously been employed to determine open cluster membership information from 3D kinematics (Gao 2013). In this work, we present the first use of DBSCAN to determine open cluster morphology from spatial distribution.

The region has also been spatially classified using the statistical method outlined by Chen et al. (2004). We have also studied the effects of extinction over the region, and analyzed the densest parts of the clusters using CMDs, from which field contamination were statistically removed with suitable representative field CMDs. Using suitably fitted isochrones, we have obtained the distance and reddening parameters for these clusters to solve the discrepancy in previous works. We have also derived a J-band luminosity function (JLF) to estimate the MF slope of the two clusters.

## 2. DATA DESCRIPTION

We obtained NIR data from the UKIDSS GPS star catalogue database (Lucas et al. 2008; data release 6). The UKIDSS project (Lawrence et al. 2007) uses the United Kingdom Infrared Telescope (UKIRT) Wide Field Camera (WFCAM; Casali et al. 2007). The photometric system is described in Hewett et al (2006), and the calibration is described in Hodgkin et al. (2009). The pipeline processing and science archive are described in Hambly et al. (2008). It delivers uniform and precise photometry in the J (1.25 μm), H (1.65 μm), and K (2.16 μm) near-infrared photometric bands. For our analysis we selected UKIDSS data from a region of size 0.865 degrees (in RA) × 0.5 degrees (in DEC), covering the previously reported cluster sizes of 6.5′ for NGC 1857 (Hoq & Clemens 2015) and 3.98′ for Czernik 20 (Zasowski et al. 2013), and the 8′ separation of their centers. To ensure the quality of data, only stars which have detections in all three photometric bands with probability of being a star greater than 0.9 have been conservatively considered for analysis.

The UKIDSS GPS is known to have saturation limits at J = 13.25 mag, H = 12.75 mag and K = 12.0 mag respectively (Lucas et al. 2008). Hence we replaced the bright stars above the UKIDSS saturation limits, within the region of analysis, with stars from the Two Micron All Sky Survey (2MASS) star catalogue database (All-Sky Data Release; Skrutskie et al. 2006). In order to obtain a homogeneous and consistent photometric dataset for our analysis, we applied photometric conversions (Hewett et al. 2006) to convert the UKIDSS dataset to the 2MASS photometric system.

For classification of possible cluster candidates, both of our clustering algorithms, mentioned in Section 3, are dependent on the number of neighboring stars to any particular star. It is understood, however, that for stars near the boundaries of the analyzed data, the number of neighboring stars is significantly reduced. Had a region smaller than the known cluster radius been considered for analysis, both our algorithms would misclassify the stars near the boundaries as non-members, resulting in the determination of an incorrect cluster morphology. Hence care was taken to ensure that the region selected for analysis was sufficiently large enough to include the known sizes of both clusters and provide enough leeway such that the classification algorithms do not misclassify stars at the boundaries.

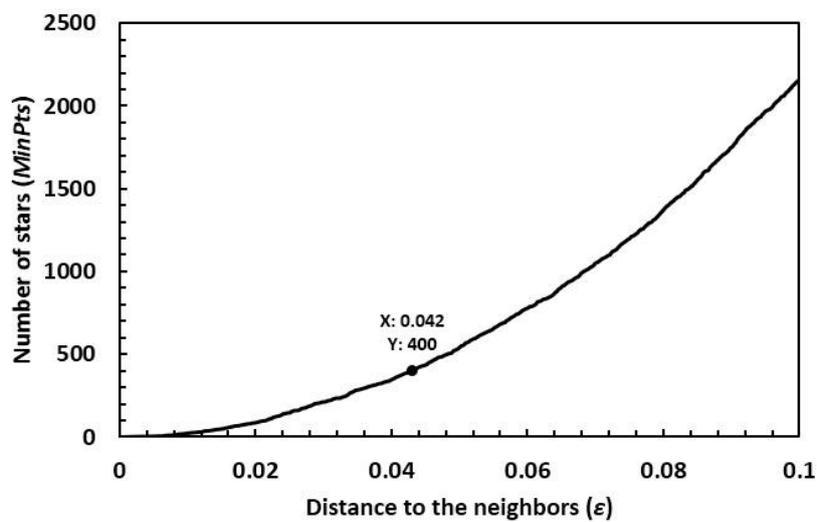

FIGURE 1: k-distance graph used to estimate the value of parameter ε for use in the DBSCAN algorithm. The value of ε and MinPts obtained have been shown.

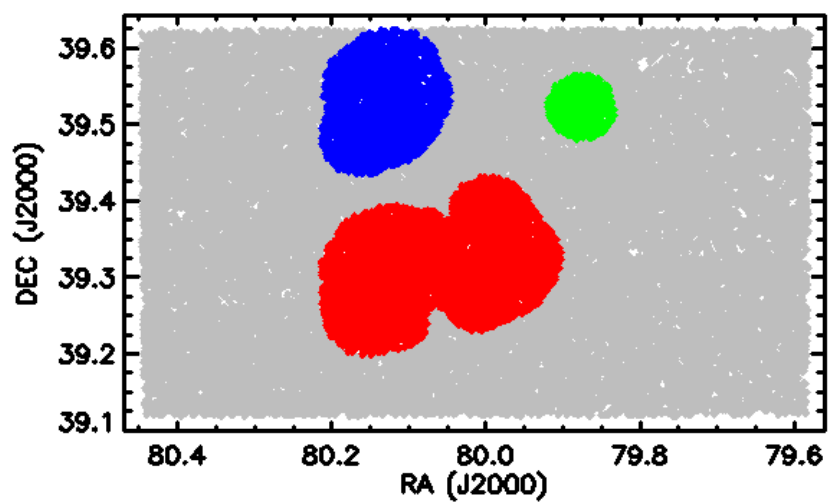

FIGURE 2: Morphology of clusters detected from the DBSCAN algorithm. The three detected clusters are shown in red, blue and green, while the stars that are not part of the cluster morphologies are depicted in grey.

## 3. CLUSTERING ALGORITHMS

### 3.1 Density-based Spatial Clustering of Applications with Noise (DBSCAN)

The DBSCAN algorithm is a density-based clustering algorithm which requires two parameters along with the data set, namely, the search radius, $\varepsilon$, and the minimum number of points, *MinPts*, in its ε-neighborhood (the circle of radius $\varepsilon$, centered on a data point) required to be considered as a cluster. The algorithm starts from an arbitrary data point in the data set. For this point, if the number of other data points in its ε-neighborhood is less than *MinPts*, the point is marked as not being part of any cluster. Otherwise a cluster is started and the process is repeated for all the points lying in the ε-neighborhood. Thus the cluster keeps growing until the terminating points, which have insufficient points in their ε-neighborhood, have been reached. At this point, the algorithm starts a new cluster by selecting a second arbitrary point and the entire process is repeated until all the data points have been exhausted. The DBSCAN algorithm does not require one to specify the number of clusters, and it can find arbitrarily shaped clusters in the parameter space. We applied DBSCAN on the entire region of the NIR spatial data (RA-DEC space) to identify spatially related points which represent cluster morphology.

The parameters $\varepsilon$ and *MinPts* have to be estimated based on the data set on which the algorithm would be running. As a lower bound, the value of *MinPts* should be greater than 3, and to account for increased density, it should be made larger as the data set gets larger. Smaller values would result in a large number of smaller clusters being identified adjacent to one another, whereas very large values would result in all clusters in the region being identified as only one large cluster. In view of the number of stars within NGC 1857 expected by Hoq & Clemens (2015), we assumed *MinPts* = 400. The value of $\varepsilon$ was estimated using a $k$-distance graph (Fig 1), where the average of the distances of every point to its $k$ (=*MinPts*) nearest neighbors are plotted. If the value chosen was too small, larger portions of the data set would not get clustered, whereas if the value chosen was too large, many of the clusters will merge together. From the k-distance graph, we obtain a value of $\varepsilon = 0.04305$ corresponding to *MinPts* = 400, and iteratively changed it to obtain $\varepsilon = 0.042$. For higher values of $\varepsilon$, the larger clusters split into separate clusters which are adjacent to each other despite clearly being the same cluster. The three clusters identified by the DBSCAN algorithms are shown in Fig 2. In section 4, we discuss the morphologies and positions of these clusters with regards to the known positions of NGC 1857 and Czernik 20.

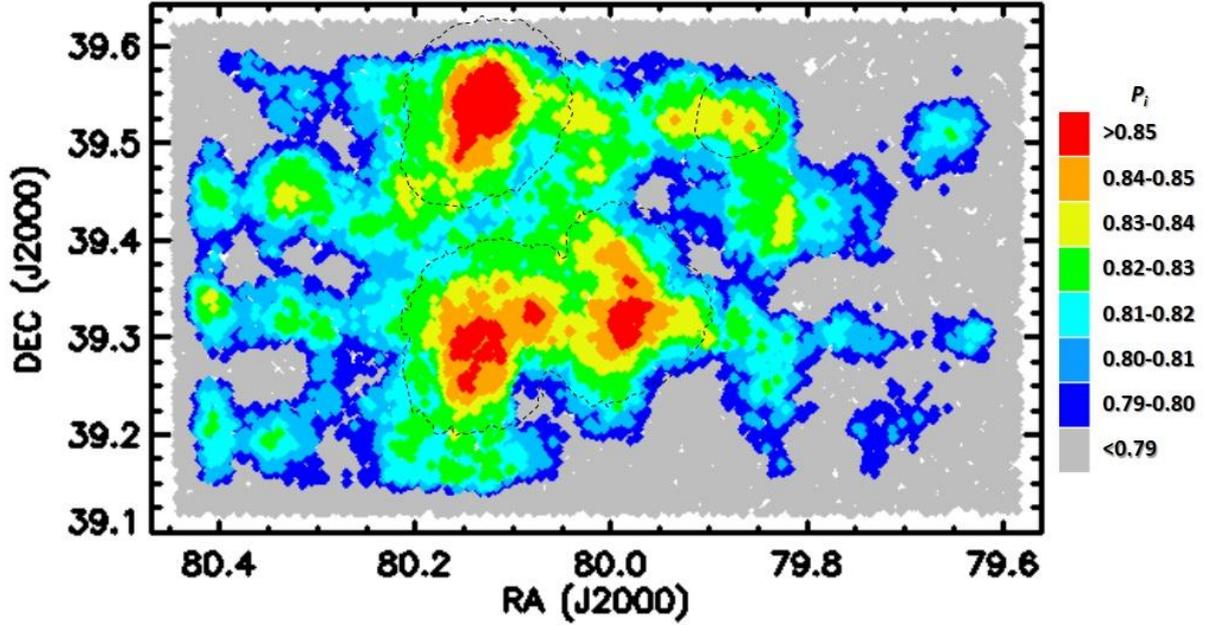

*Figure 3: Representation of spatial surface density of stars, depicting the morphology of the clusters as obtained from the statistical membership determination algorithm. The clustering parameter values are shown in the color bar. The dotted lines show the outline of the cluster morphology obtained from DBSCAN.*

### 3.2 Statistical Membership Determination

The probabilistic star counting technique has been used to statistically determine the morphologies of NGC 1857 and Czernik 20. A clustering parameter for each star, which behaves like a quantitative measure of cluster membership, is defined as $P_i = (N_t - N_f)/N_t$, where $N_t$ is the total number of neighbouring stars within a specified angular size centered on the star, termed the neighborhood aperture $r_p$, and $N_f$ is the average number of field stars within the same aperture. We have assumed $N_f = 50$ (Chen et al. 2004) to calculate $r_p$ such that $P_i \geq 0$ for all the stars in our dataset, including field stars. This implied that a non-zero $P_i$ had to be chosen as the lower cut-off value for possible cluster members, since a considerably large number of stars had been assigned a relatively high $P_i$ value.

We thus obtained $r_p = 0.037$ and calculated $P_i$ accordingly for each star in our dataset. Fig 3 shows the spatial distribution of all stars in our dataset. Stars with different $P_i$ are shown with different colours, as per the color index. The distribution of $P_i$ also represents the surface number density of stars, with those shown in red having the highest surface number density. In view of the surface stellar density thus obtained, we consider the stars shown in yellow, with $P_i > 0.83$ as the cut-off for cluster boundaries. The morphologies of the open clusters can be visualized from the region bordered by the stars marked in yellow in Fig 3 and are consistent with that of NGC 1857 and Czernik 20, studied by Zasowski et al. (2013). It is to be noted that other cut-off values around 0.83 still resulted in a similar morphology.

## 3.3 Comparison of Clustering Algorithms

From the morphologies of the clusters obtained by implementing the two clustering algorithms, it is quite evident that they agree well with each other even though the cluster morphologies obtained from DBSCAN are larger. Czernik 20 seems to exhibit a nearly circular morphology whereas NGC 1857 seems to have an irregular morphology with two dense cores. Contrary to the claims by Sujatha et al. (2006), we find that the two clusters are distinctly separate. Based solely on spatial distribution information, both the clusters determine possible cluster morphologies by employing suitable variations of nearest neighbor statistics. Yet they differ somewhat in the morphologies that they represent.

While DBSCAN, by virtue of being a hard clustering algorithm, classifies stars as being either a possible cluster candidate or not, the statistical membership determination algorithm assigns a clustering parameter to each star and the specific cut-off value to be used to distinguish a possible cluster candidate from a non-candidate rested with us. However, since the variation in clustering parameter was representative of the surface number density of stars in any analyzed region, we utilized it to determine the densest regions in any cluster. Identification of the densest regions within the cluster morphologies would not have been possible from DBSCAN.

We also note that the representation of surface density by the clustering parameter is susceptible to the relative density of different clusters in the region being analyzed. For example, if there is a cluster with very high surface density in the region being analyzed, then other less dense clusters in the region don't quite stand out and may even be disregarded depending on the $P_i$ cutoff. The density of other clusters in the region, however, has little bearing on the DBSCAN algorithm's ability to identify clusters in a region. In Section 4.5, we have investigated whether the additional cluster identified by DBSCAN is consistent with being a star cluster or not. We also point out that the parameter ε used in DBSCAN, and the neighborhood aperture used in Statistical Membership Determination, appear to be identical in terms of their physical interpretation. They both refer to the radius of the circle centered on a point within which other points on the RA-DEC space are considered neighbors. The slight variation in values is because of the difference in the methods by which they are obtained.

## 4. ANALYSIS

The surface number density of the star clusters may also be affected by the spatial variation in extinction. We therefore investigated the extinction variation before characterizing the cluster morphology. Additionally, the spatially deduced possible cluster candidates identified by the two clustering algorithms still suffered from field contamination. We exercised a statistical cleaning of the CMDs (similar to Pandey et al. 2008) for the densest parts of the clusters to remove field contamination, and obtained representative cleaned CMDs of the clusters which were further used to obtain important parameters, like distance and extinction. Those sources with photometric uncertainties less than 0.1 mag in all three bands have been used for this purpose.

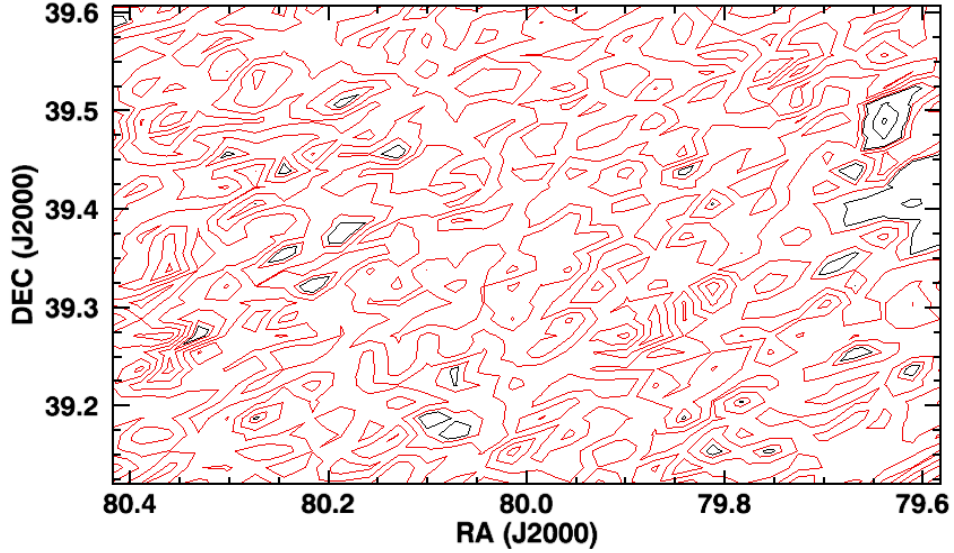

*Figure 4: $A_K$ Extinction Map of the analyzed region. $A_K$ values are estimated from the $(H - K)$ colors. The contour levels are for $A_K$ values 0.02–0.18 (red) and 0.2–0.4 mag (black) respectively.*

### 4.1 Extinction Map

We estimated the K-band extinction ($A_K$) towards the entire region considered for analysis using the 2MASS and UKIDSS data. We measured $A_K$ using the $(H - K)$ colors of the stars. For every star, the observed (H − K) value was converted into $A_K$ using the reddening law given by Flaherty et al. (2007), i.e., $A_K = 1.82 \times (H - K)_{obs} - (H - K)_0$, where $(H - K)_0$ is the average intrinsic color of stars, which was assumed to be 0.2 (Allen et al. 2008; Gutermuth et al. 2009). The extinction map of the region of analysis (Fig 4) was obtained by dividing it into 900 grids of equal area and the mean $A_K$ in each grid was calculated along with the standard deviation for each star. It is to be noted that we rejected those sources deviating above 3σ to calculate the mean value of $A_K$ in each grid. The mean $A_K$ for the whole region is ∼0.13, while $\varDelta A_K$ is found to be ∼0.48. However, we noticed from Fig 4 that the region where with a supposed enhancement of surface density indicating a presence of a cluster, from both the clustering algorithms (Fig 2, 3), was characterized by $\varDelta A_K < 0.2$, as evident from the contours shown in red. This implied that variation in extinction is not responsible for the enhanced surface density and there is indeed an increased stellar density in those regions.

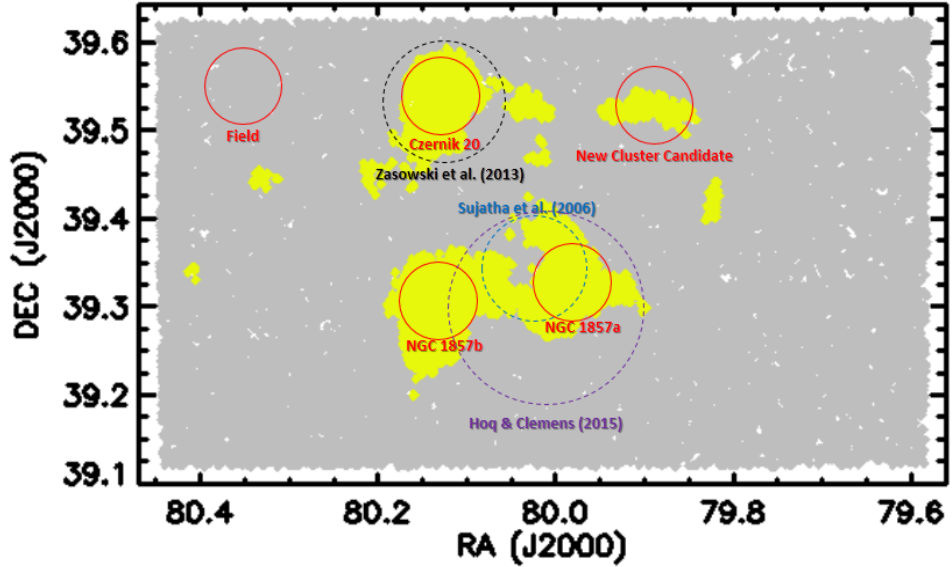

*Figure 5: The various regions analyzed in this work (red) and in previous works (other colors) have been plotted against a background of the cluster morphology obtained from statistical membership determination.*

**4.2 Field Decontamination of Color Magnitude Diagrams**

For the regions, radius 2.5′ each, with the highest stellar density, as shown in Fig. 5, we obtained J-K vs J CMDs for the 2MASS and UKIDSS data with photometric uncertainties less than 0.1 in all three bands. The analyzed regions are termed as Czernik 20 (centered around RA = 80.130, DEC = 39.540), NGC 1857a (centered around RA = 79.985, DEC = 39.325), and NGC 1857b (centered around RA = 80.135, DEC = 39.305). We also marked the regions analyzed by previous works (Hoq & Clemens 2015, Zasowski et al. 2013, Sujatha et al. 2006) in Fig 5. To statistically remove the field contamination from these regions, we chose a nearby comparison field of radius 2.5′, shown in Fig 5, having similar $A_K$ variation as the cluster.

The J-K vs J CMD for the field (Fig 6) was compared with those of our three target cluster CMDs (Fig 7a). For every star found in the field CMD, a star was eliminated from the target CMD if it laid within a box, of size 0.56 mag in J-K and 0.2 in J, centered on the field star. We chose this particular box size for cleaning because it not only accounted for the photometric uncertainty in magnitudes, but was also the one deemed most suitable after trying out both smaller boxes, which proved ineffective in cleaning sufficient field-star counterparts, and larger ones, which cleaned out possible cluster members.

To establish that the chosen field was truly random and representative of the field region around the target clusters, another field, having the same galactic latitude so as to minimize any possible density gradient of field stars, was first cleaned using our field decontamination method. Upon finding a sufficiently cleaned CMD of the second field and corroborating the truly random nature of our chosen field from the cleaned CMD, we proceeded to decontaminate the cluster regions.

The cleaned CMDs of the target cluster regions are shown in Fig 7b. In Fig 7c, we also show those stars in the field which did not eliminate any cluster counterpart, to demonstrate the quality of our statistical cleaning from the sparsely distributed field stars which are uncorrelated to the isochrone of best fit (Section 4.3). Despite this procedure, we cannot claim that the field has been perfectly cleaned since a few stars in the cleaned CMD still appear to be unrelated to the cluster.

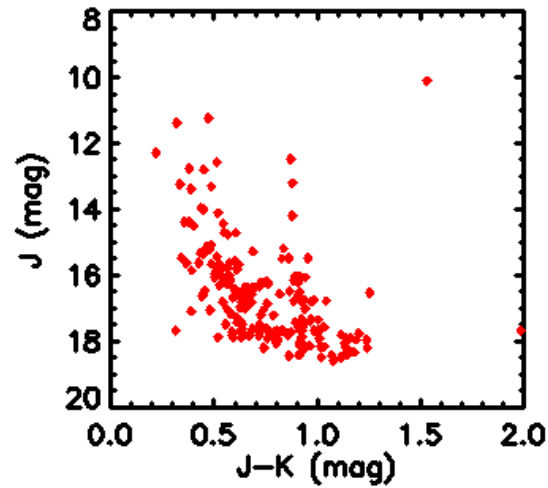

*Figure 6: CMD for the field used for decontamination.*

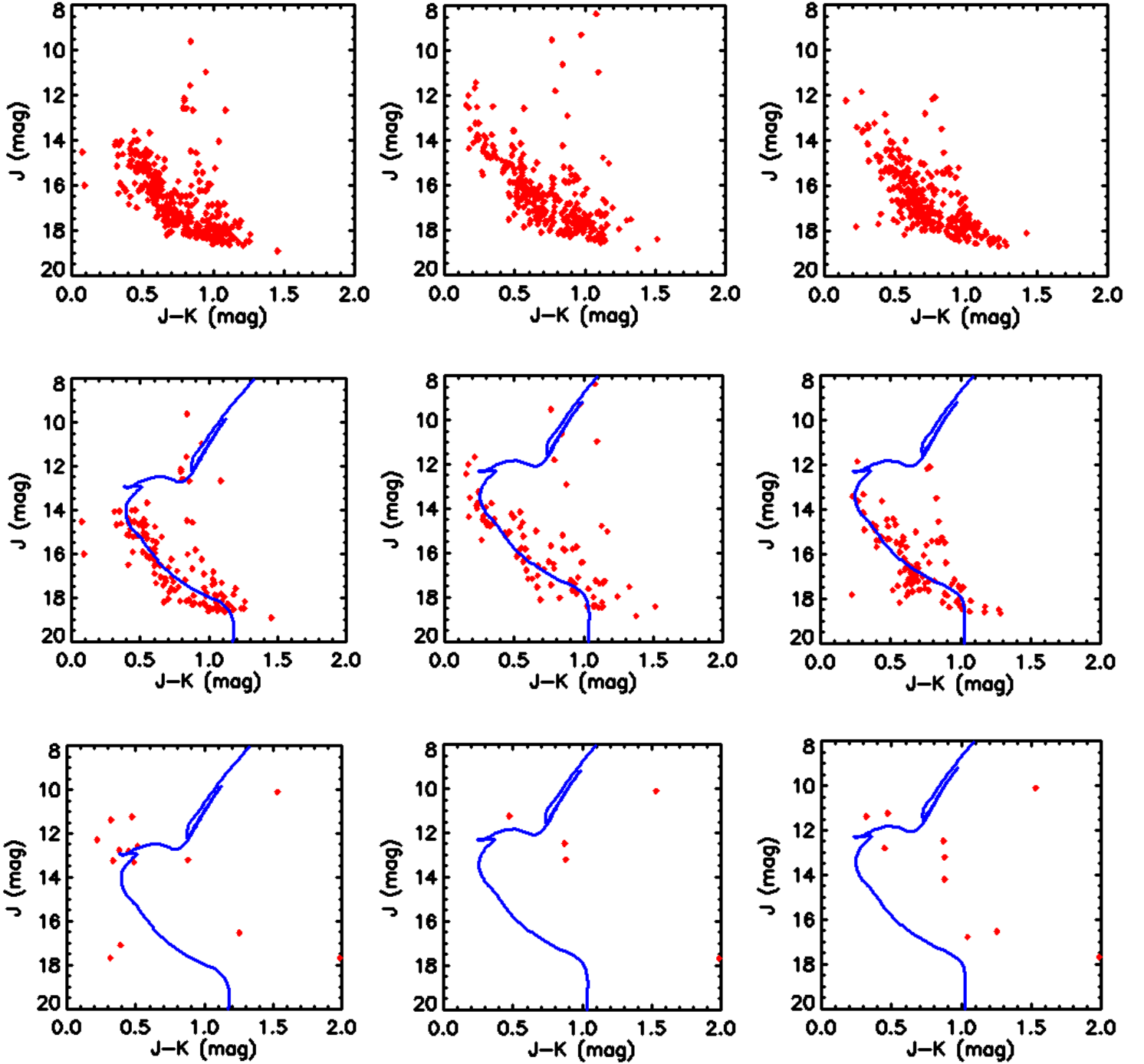

*FIGURE 7: From left to right, the three columns show the regions studied- Czernik 20, NGC 1857a and NGC 1857b, respectively. a) The top panel shows the CMDs of the cluster regions analyzed. b) The center panel shows the CMDs of the cluster regions obtained after field decontamination, along with the isochrone of best fit. c) The lower panel shows the field stars which did not eliminate a counterpart in the cluster CMD, shown to be uncorrelated with the isochrone of best fit.*

### 4.3 Isochrone Fitting

By comparing the cleaned CMDs with the PARSEC stellar evolution isochrones (Bressan et al. 2012), in particular including the improvement on very low-mass stars (Chen et al. 2014), we inferred the stellar properties. The age was fixed as 1 Gyr and metallicity as $Z_\odot$ (which is equal to 0.0152 for PARSEC isochrones), consistent with those considered by previous studies (Zasowski et al. 2013; Kharchenko et al. 2013). We considered $A_J = 1.62 \times E(J-K)$ (Tokunaga 2000), and varied *E(J-K)* and the distance to fit the isochrones to the CMDs of the cluster regions.

For each star identified in our cleaned cluster CMD, a circle of unit radius was obtained in the J vs 5*(J-K) parameter space. The scaling up of the color axis by a factor of 5, accounted for the difference in the range of values pertaining to the J axis (~7.5 mag) and J-K axis (~1.5 mag). The separation from the point on the reddened isochrone, within the circle, nearest to the position of the star on this parameter space, was computed as $d_{star}$. The mean $d_{star}$ of those stars was computed as $d_{iso}$. We minimized this value to obtain the isochrone of best fit for the cleaned cluster CMD (Fig 7b), from which the distance and reddening were inferred. The *E(J-K)* and distance to the clusters derived from our fitted isochrones, along with those from previous works, are presented in Table 1.

| Region analyzed | This Work | | Zasowski et al. (2013) | | Kharchenko et al. (2013) | | Hoq & Clemens (2015) | |
|---|---|---|---|---|---|---|---|---|
| | *E(J-K)* (mag) | Distance (pc) | *E(J-K)* (mag) | Distance (pc) | *E(J-K)* (mag) | Distance (pc) | *E(J-K)* (mag) | Distance (pc) |
| **Czernik 20** | 0.33 | 2900 | 0.35 | 3100 | 0.288 | 2000 | - | - |
| **NGC 1857 (whole)** | - | - | 0.07 | 1400 | 0.24 | 3299 | 0.31 | 2470 |
| **NGC 1857a** | 0.19 | 2400 | - | - | - | - | - | - |
| **NGC 1857b** | 0.18 | 2400 | - | - | - | - | - | - |

*Table 1: Parameters of analyzed regions derived from the isochrone of best fit, and those from previous studies.*

### 4.4 Luminosity Functions and Mass Functions

We constructed apparent J-band Luminosity functions (JLFs) for the stars present in the three regions after decontamination (Fig 8). It is to be noted that we only included those stars in the JLFs which had magnitudes below the turn off regions for the respective isochrones. The mass function (MF) for the stars present in the three regions, after decontamination, were obtained using the JLFs and the mass information from the Parsec isochrones (Chabrier 2001; Salpeter 1955), incorporating the distance and extinction information previously obtained. They have been shown in Fig 9. The MF slope (Bastian et al. 2010) was derived by using the relation $\log\left(\frac{dN}{dM}\right) = -(1+x)\log(M) + constant$, where *dN* represents the number of stars in a mass bin *dM* with central mass *M* and *x* is the slope of the MF.

The MF of the densest part of Czernik 20 has a slope of 1.03 in the mass range 0.65-2.11 $M_\odot$, whereas those for NGC 1857a and NGC 1857b have slopes of 1.15 and 1.38 respectively in the mass range 0.55-2.05 $M_\odot$. They resemble the slope of 1.35 obtained by Salpeter (1955) for field stars in the solar neighborhood, and also agree with that obtained by Sujatha et al. (2006) for NGC 1857.

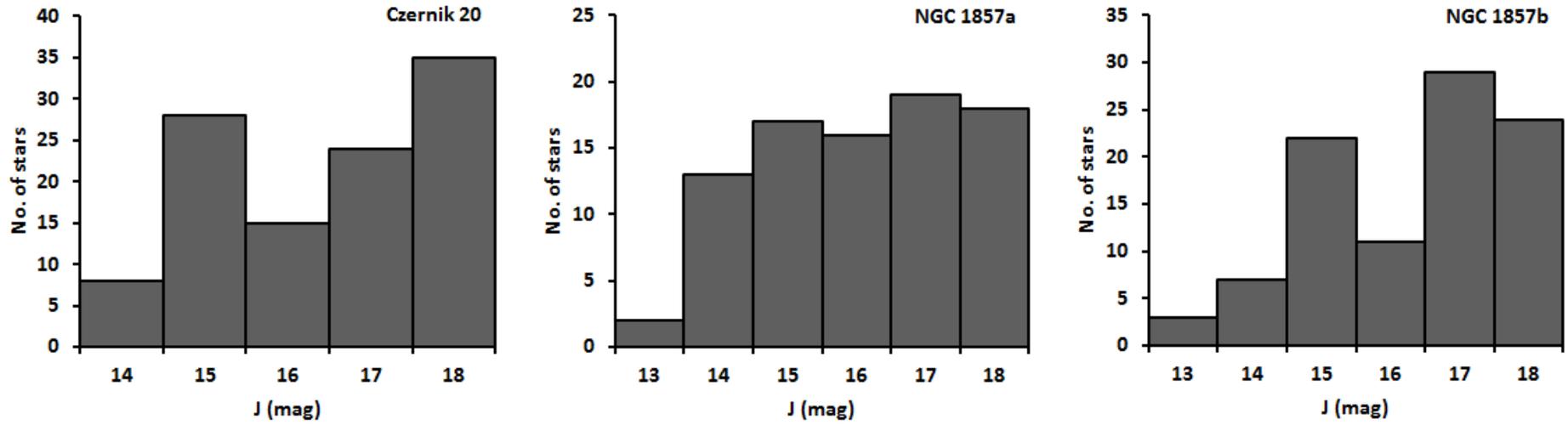

*FIGURE 8: The J-band luminosity functions for Czernik 20 (left), NGC1857a (center) and NGC 1857b (right).*

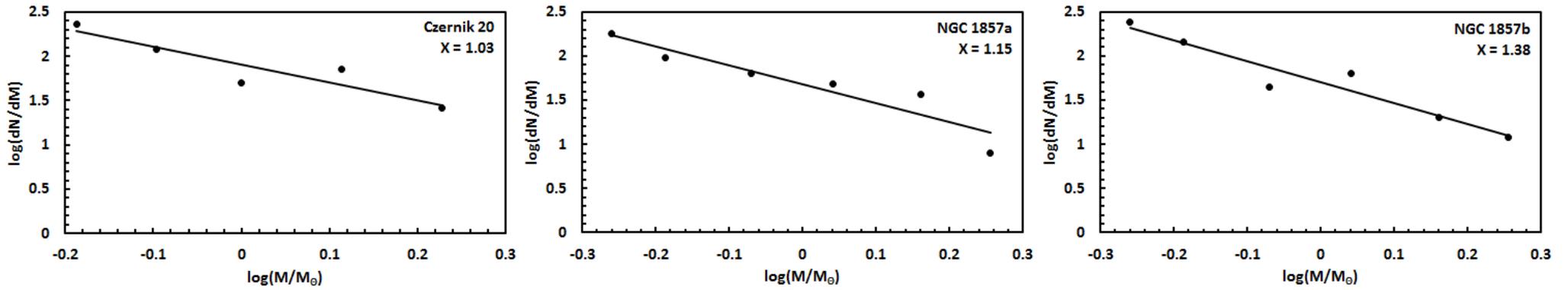

*FIGURE 9: The mass functions for Czernik 20 (left), NGC1857a (center) and NGC 1857b (right).*

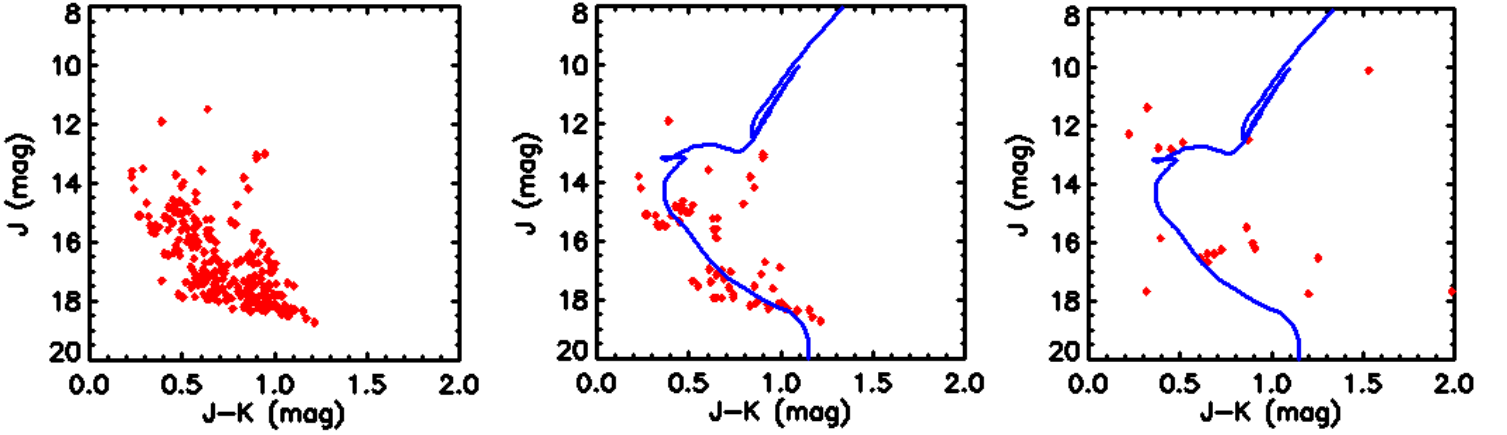

*FIGURE 10: CMD of the new cluster candidate (left). CMDs of the new cluster candidate obtained after field decontamination, along with the isochrone of best fit (center). The field stars which did not eliminate a counterpart in the cluster CMD, shown to be uncorrelated with the isochrone of best fit.*

### 4.5: The New Cluster Candidate

The DBSCAN clustering algorithm found a third cluster, along with those around the previously studied regions of NGC 1857 and Czernik 20, centered nearly around RA=79.89 and DEC=39.532. From the statistical membership determination algorithm, we found an enhanced surface stellar density around this region. We also noted (Fig 4) that the variation in $A_K$ around this region is less than 0.2. This led us to investigate a 2.5′ region around the aforementioned center to obtain a CMD for the region and we subsequently statistically cleaned the field contamination to obtain the cleaned CMD (Fig 10). It clearly shows a main sequence consistent with that of star clusters. We fitted an isochrone of age 1 Gyr and solar metallicity, to obtain a distance of 3300 pc and E(J-K) = 0.3. We thus find an over-density having parameters consistent with that of a star cluster, which may be related to Czernik 20 (owing to the comparable distance and reddening) or may be a separate star cluster.

### 5. DISCUSSION

We have presented the first use of the DBSCAN algorithm for finding morphologies of open clusters from spatial distribution. It is to be noted that the ability of DBSCAN to identify clusters in parameter space, independent of the densities of other clusters in the same space is what led to the identification of the additional cluster candidate. This implies that the DBSCAN algorithm can be effective in identifying cluster candidates in large all-sky surveys. We have also corroborated the effectiveness of the statistical membership determination algorithm in finding open cluster morphologies and representing surface stellar densities.

Through our analysis, we re-establish the conclusion reached by Zasowski et al. (2013), that Czernik 20 and NGC 1857 are indeed separate clusters, and belie the conclusion reached by Sujatha et al. (2006). Table 1 shows that the distance and reddening derived by us for Czernik 20 correspond well with those derived by Zasowski et al (2013) but don't agree with those derived by Kharchenko et al. (2013). For the case of NGC 1857, the distance and reddening derived by us correspond well with the distance derived by Hoq & Clemens (2015) but not with their higher reddening. Both of our derived parameters don't agree with those obtained by Zasowski et al. (2013) and by Kharchenko et al. (2013). The difference in our derived parameters was probably caused due to the difference in regions of NGC 1857 analyzed by us and previous works (Fig 5). In particular, none of the previous works had studied the region around NGC 1857b. We also note that Hoq & Clemens (2015) had fitted an isochrone with Z=0.005 which, as opposed to the Z= $Z_\odot$ isochrone fitted by us, would certainly require a higher reddening to fit the same sequence. Thus, the region selected for analysis in conjunction with UKIDSS being a deeper survey than 2MASS leads us to conclude that the parameters derived by us for NGC 1857 are probably more reliable than those obtained from previous works.

The morphology of NGC 1857, characterized by the presence of two dense core regions, is probably a result of an interesting dynamical evolution and will be the subject of detailed analysis in the future. We derive mass function slopes for Czernik 20 and NGC 1857, which are in agreement with the value of 1.35 obtained by Salpeter 1955 for field stars in the solar neighborhood. If we were to assume that these values are consistent even for those stars which are too faint to be identified by reliable UKIDSS photometry, extrapolation of the MFs down to masses of 0.1 $M_\odot$ would indicate the presence of ~4700, ~3000 and ~5500 additional stars for Czernik 20, NGC 1857a and NGC 1857b respectively. However, the mass function slope is expected to gradually become less steep as we approach 0.1 $M_\odot$ (Bastian et al. 2010), indicating that actually far fewer stars have not been observed. It is also worth mentioning that the new cluster candidate identified by us shows a clear main sequence, which has been fit to identify the distance and reddening but we have not derived an age due to the lack of a distinct red clump.

**Acknowledgements:**

We thank Dr. Wen-Ping Chen, Director, Graduate Institute of Astronomy, National Central University, Taiwan, and the anonymous referee for their valuable comments and suggestions which greatly improved the quality of the paper. This publication makes use of data from The UKIRT Infrared Deep Sky Survey (UKIDSS) which is a next generation near-IR sky survey using the wide field camera (WFCAM) on the United Kingdom Infrared Telescope on Mauna Kea in Hawaii. This work also makes use of data products from the Two-Micron All-Sky Survey (2MASS), which is a joint project of the University of Massachusetts and the Infrared Processing and Analysis Center/ California Institute of Technology, funded by the National Aeronautics and Space Administration and the National Science Foundation. This research has also made use of the VizieR catalogue access tool, CDS, Strasbourg, France.